\documentclass[a4paper,fleqn]{cas-dc}
\usepackage[numbers]{natbib}
\usepackage{orcidlink}
\makeatletter
\let\@ORCID\@empty

\makeatother

\begin{document}
\let\WriteBookmarks\relax
\def\floatpagepagefraction{1}
\def\textpagefraction{.001}
\shortauthors{Zhou et~al.}
\shorttitle{Med-URWKV$\dagger$: Toward Enhanced Pretrained Pure VRWKV Models for Medical Image Segmentation}

\title [mode = title]{Med-URWKV$\dagger$: Toward Enhanced Pretrained Pure VRWKV Models for Medical Image Segmentation}                      

\author[1,2]{Zhenhuan Zhou \orcidlink{0009-0000-2187-7184}}[style=chinese]

\author[3]{Yining Li \orcidlink{0009-0006-2564-7486}}[style=chinese]

\author[1,2]{Yanlin Wu \orcidlink{0000-0002-2087-275X}}[style=chinese]

\author[4,5]{Haohan Zou \orcidlink{0000-0003-0588-9753}}[style=chinese]

\author[4,5]{Yan Wang* \orcidlink{0000-0002-1257-6635}}[style=chinese]

\author[1,6]{Tao Li* \orcidlink{0000-0002-1273-0487}}[style=chinese]

\affiliation[1]{organization={College of Computer Science, Nankai University},
            city={Tianjin},
            postcode={300350}, 
            country={China}}
            
\affiliation[2]{organization={Key Laboratory of Data and Intelligent System Security, Ministry of Education},
            country={China}}

\affiliation[3]{organization={School of Medicine, Nankai University},
            city={Tianjin},
            postcode={300350}, 
            country={China}}

\affiliation[4]{
            organization={Nankai University Eye institute, Nankai University},
            city={Tianjin},
            country={China}
            }

\affiliation[5]{
            organization={Tianjn Eye Hospital},
            city={Tianjin},
            country={China}
            }

\affiliation[6]{organization={Haihe Lab of ITAI},
            city={Tianjin},
            postcode={300459}, 
            country={China}}

\cortext[cor1]{Co-corresponding Authors: Yan Wang (wangyan7143@vip.sina.com) and Tao Li (litao@nankai.edu.cn)}

\nonumnote{Zhenhuan Zhou and Yining Li contribute equally.}

\begin{abstract}
Medical image segmentation is a fundamental task in computer-aided diagnosis and treatment. Existing approaches based on CNNs, ViTs, Mamba, and hybrid models still suffer from limitations such as restricted receptive fields, high computational cost, or insufficient accuracy. Recently, Vision Receptive-field Weighted Key-Value (VRWKV) models have emerged as a promising alternative,delivering strong long-range dependency modeling for visual tasks. However, current studies on VRWKV-based medical image segmentation mainly focus on hybrid architectures trained from scratch, while the potential of large-scale pretrained pure VRWKV models remains unexplored. In this work, we systematically investigate the effectiveness of pure VRWKV architectures for medical image segmentation. We construct Med-URWKV-T and Med-URWKV-S by reusing pretrained VRWKV encoders at different scales and pairing them with pure VRWKV decoders, enabling a comprehensive evaluation of pretrained pure VRWKV models in this domain. To further enhance performance, we propose two VRWKV-compatible modules: a Frequency-Aware Wavelet Attention (FAWA) module, which exploits wavelet transforms to capture edge details and structural characteristics, and a Multi-Scale Channel Fusion (MSCF) module, which integrates multi-scale features to strengthen informative channel representations. By incorporating them into Med-URWKV-T, we obtain the enhanced model Med-URWKV$\dagger$. Extensive experiments on five medical image segmentation datasets demonstrate that Med-URWKV achieves performance comparable to or superior to state-of-the-art methods and carefully designed hybrid VRWKV architectures. Moreover, Med-URWKV$\dagger$ further improves segmentation accuracy, surpassing Med-URWKV-S while using only half of its parameter count, and achieves the highest average Dice similarity coefficient of 88.00\%. The codes will be released.
\end{abstract}

\begin{highlights}
\item A Med-URWKV model series is constructed to systematically explore the potential of large-scale pretrained pure VRWKV architectures for medical image segmentation.

\item Experimental results demonstrate that pretrained pure VRWKV models achieve strong segmentation performance, while revealing a clear performance bottleneck when model capacity is simply increased.

\item A Frequency-Aware Wavelet Attention (FAWA) module and a Multi-Scale Channel Fusion (MSCF) module are proposed to enhance frequency-domain modeling and multi-scale feature representation.

\item The resulting Med-URWKV+ consistently outperforms 17 representative deep learning methods across five public datasets, achieving a superior balance between parameter complexity and segmentation accuracy.

\item Ablation studies verify the effectiveness of large-scale pretraining and the contributions of the FAWA and MSCF modules in improving pure VRWKV-based segmentation models.

\end{highlights}


\maketitle

\section{Introduction}
Medical segmentation serves as a critical cornerstone for computer-aided diagnosis (CAD) and treatment planning, it is also one of the most significant application directions of deep learning (DL) in the medical field. The automatic segmentation of anatomical structures and pathological lesions can not only significantly reduce the workload of clinicians and mitigate inter-observer variability, but also provide an objective and quantitative basis for the optimization and formulation of individualized treatment plans. Ultimately, this enables the achievement of goals such as alleviating patients' discomfort during diagnosis and treatment, as well as improving clinical prognostic outcomes.

\begin{figure}[pos=htbp]
\includegraphics[width=\columnwidth]{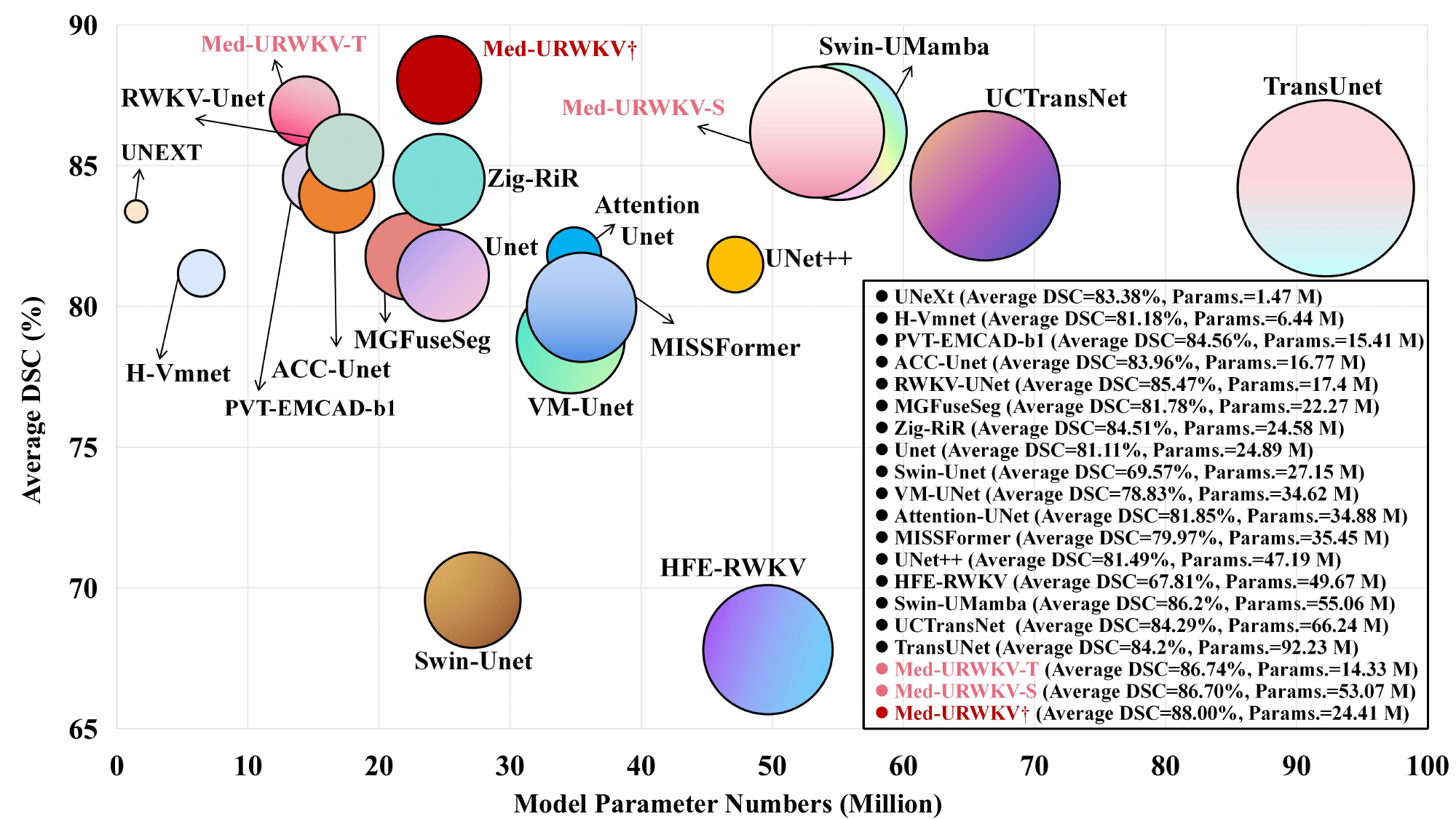}
\caption{Comparison of Med-URWKV$\dagger$ and other methods in terms of average DSC versus model parameter count. The area of each circle is proportional to the number of model parameters. As shown, the improved Med-URWKV$\dagger$ achieves a better balance between average segmentation performance and model complexity.} 
\label{fig1}
\end{figure}

With the advent of convolutional neural networks (CNNs), DL has been widely adopted for medical image segmentation. U-Net \cite{ronneberger2015u} and its variants \cite{he2022progressive, rahman2024emcad, farshad2022net, li2023mixunet} have become representative models due to their strong local feature extraction capability. However, limited receptive fields restrict CNNs in tasks requiring global context modeling or long-range dependency learning. The introduction of self-attention \cite{vaswani2017attention} and Vision Transformers (ViT) \cite{dosovitskiy2020image} alleviates this limitation, but their quadratic computational complexity leads to high computational and memory costs for high-resolution medical images \cite{zhai2021attention}, hindering clinical deployment. More recently, Mamba \cite{gu2023mamba} has emerged as a promising alternative, and several Mamba-based models \cite{liu2024vmamba, 10821761, gong2025nnmamba} have been successfully applied to medical image segmentation. Although Mamba offers linear computational complexity and improved efficiency, this advantage is often accompanied by a trade-off in segmentation accuracy \cite{chen2025zig}. The recently proposed architecture, Receptance Weighted Key Value (RWKV) \cite{peng2023rwkv}, has attracted increasing attention. Benefiting from its linear computational complexity, strong capability in modeling long-range dependencies, and a parallelizable training mechanism, RWKV has quickly garnered interest from researchers across various fields \cite{11039697}, \cite{xu2025urwkv}, \cite{zhou2025personalizable}. Duan et al. extended RWKV to the vision domain by introducing Vision-RWKV (VRWKV) \cite{duan2024vision}, which demonstrates outstanding performance on a variety of visual tasks \cite{yuan2024mamba}, \cite{zhang2025out}. In the field of medical image segmentation, VRWKV has also received initial attention \cite{chen2025zig, jiang2025rwkv, ye2025hfe}.

However, based on our observations, existing applications of VRWKV in medical image segmentation still exhibit several limitations. First, most approaches rely on hybrid architectures trained from scratch, which fail to fully exploit the true potential of large-scale pretrained pure VRWKV models when transferred to medical image segmentation tasks. Second, although RWKV models excel at modeling long-range dependencies, their ability to integrate global and local features, as well as their robustness to spatial continuity, remains insufficient \cite{chen2025zig}. Moreover, despite the proven importance of frequency-domain learning in medical image segmentation \cite{zhou2024spatial, zhou2023xnet}, the capability and effectiveness of pure VRWKV models in learning frequency-domain representations have yet to be thoroughly explored.

Therefore, this work is guided by two main research directions. First, we explore the transferability and true performance of large-scale pretrained pure VRWKV models at different scales in the field of medical image segmentation. Second, we aim to address the inherent limitations of pure VRWKV architectures by designing targeted improvement strategies, further enhancing their segmentation performance without introducing other architectural paradigms. For the former, inspired by prior work \cite{liu2024swin}, we construct two pure VRWKV architectures at different scales by pairing ImageNet-pretrained \cite{deng2009imagenet} VRWKV encoders with pure VRWKV decoders, resulting in the Tiny and Small versions of Med-URWKV. For the latter, we specifically design a Frequency-Aware Wavelet Attention (FAWA) module and a Multi-Scale Channel Fusion (MSCF) mechanism to enhance frequency-domain perception, improve the modeling of high-frequency details and boundaries, and strengthen multi-scale feature aggregation. By integrating these components into Med-URWKV, we obtain the further optimized Med-URWKV$\dagger$. Extensive experiments on five public datasets demonstrate that Med-URWKV outperforms previous models based on CNNs, ViTs, and Mamba. Moreover, compared with carefully designed hybrid VRWKV architectures, it achieves comparable or even superior performance. The improved Med-URWKV$\dagger$ further enhances segmentation accuracy, achieving better performance than Med-URWKV-S while using less than half of its parameters.

In summary, the main contributions of this paper can be summarized as follows:
\begin{itemize}
    \item By reusing pretrained VRWKV encoders, we construct the Med-URWKV series to investigate the transferability and performance of pure VRWKV models in medical image segmentation tasks, as well as their performance variations under different parameter scales.
    
    \item Based on Med-URWKV, we further propose Med-URWKV$\dagger$, which incorporates two core modules: the FAWA module and the MSCF module. The former improves the precision of feature detail representation by introducing frequency-domain information, while the latter enhances feature aggregation through multi-scale fusion.
    
    \item Extensive experiments are conducted on five datasets, with comparisons against 17 DL models of diverse architectures. The results show that Med-URWKV achieves comparable or superior performance to existing mainstream networks and previously proposed hybrid VRWKV models, while the optimized Med-URWKV$\dagger$ further improves segmentation performance. Ablation studies also confirm the effectiveness of the pretraining strategy and the proposed modules.
\end{itemize}

\section{Related Works}
\label{Related Works}

\subsection{Medical Image Segmentation}
\begin{figure*}[pos=htbp]
\includegraphics[width=\textwidth,trim=2.8cm 15cm 2.4cm 11.3cm,clip]{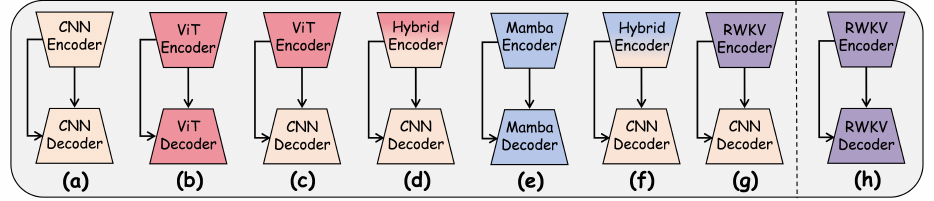}
\caption{Conceptual illustrations of representative medical image segmentation architectures: (a) pure CNN; (b) pure ViT; (c) ViT encoder with CNN decoder; (d) hybrid CNN–ViT encoder with CNN decoder; (e) pure Mamba; (f) hybrid CNN–Mamba encoder with CNN decoder; (g) RWKV-based encoder with CNN decoder; and (h) pure RWKV (ours).
} \label{fig2}
\end{figure*}

In recent years, DL has been widely applied to medical image segmentation, leading to the development of diverse model architectures, as illustrated in Fig. \ref{fig2}. With the advancement of CNNs, numerous segmentation methods represented by U-Net \cite{ronneberger2015u} and its variants \cite{li2018h}, \cite{bui20173d}, \cite{milletari2016v} have been proposed, all adopting pure CNN-based encoders and decoders (Fig. \ref{fig2}(a)). Representative extensions include 3D U-Net \cite{cciccek20163d}, which enables volumetric segmentation under sparse annotations, and Attention U-Net \cite{oktay2018attention}, which enhances feature representation through attention gates. Subsequent works further improve CNN-based frameworks by introducing adaptive attention mechanisms \cite{chen2022aau}, multi-context information fusion \cite{li2023can}, and efficient multi-scale attention decoders \cite{rahman2024emcad}. Despite their success, CNN-based methods remain constrained by limited receptive fields and insufficient global feature modeling capability, which are critical for complex medical image segmentation tasks.

With the introduction of ViTs \cite{dosovitskiy2020image}, ViT-based models have been widely applied to medical image segmentation \cite{huang2023polyp2former}, \cite{wang2025set}, \cite{10822736}. Pure ViT-based architectures (Fig. \ref{fig2} (b)), such as Swin-Unet \cite{cao2022swin} and MissFormer \cite{huang2021missformer}, enhance global context modeling while improving local feature representation through structural refinements. To combine the strengths of CNNs and ViTs, hybrid architectures have been proposed, including ViT encoders with CNN decoders (Fig. \ref{fig2} (c)) and hybrid encoder designs (Fig. \ref{fig2} (d)). Some representative methods include TransUNet \cite{chen2021transunet}, UNETR \cite{hatamizadeh2022unetr}, and H2Former \cite{he2023h2former}, which jointly leverage local feature extraction and global dependency modeling. Nevertheless, the quadratic complexity of self-attention remains a major limitation, restricting efficiency and clinical deployment.

The introduction of Mamba \cite{gu2023mamba} has provided a new direction for medical image segmentation. Based on state space models (SSM), Mamba captures long-range dependencies with linear computational complexity, offering improved efficiency and lower memory consumption compared to ViT. As a result, numerous VMamba-based models (Fig. \ref{fig2} (e) and (f)) have been applied to medical image segmentation \cite{liu2024vmamba}, \cite{10821714}, \cite{10821761}. Representative methods include Lightm-UNet \cite{liao2024lightm}, which adopts a pure Mamba-based architecture to extract deep semantic features, nnMamba \cite{gong2025nnmamba}, which integrates SSM into convolutional residual blocks to jointly model local and long-range dependencies, and a large-kernel VMamba U-shaped network \cite{wang2024large}, which expands the effective receptive field to enhance spatial representation learning. However, existing studies indicate that the improved efficiency of Mamba often comes at the cost of reduced segmentation accuracy compared to Transformer-based models \cite{chen2025zig}.

\subsection{VRWKV in Medical Image Analysis}
The introduction of RWKV \cite{peng2023rwkv} presents another promising solution, and a growing number of researchers have begun to explore its applications in medical image analysis. For example, Wang et al. \cite{wang2025feat} proposed the FEAT architecture, which leverages the attention mechanism of RWKV through a spatiotemporal-channel attention module to achieve high-quality medical video generation. He et al. \cite{he2025rwkvmatch} extended RWKV to the 3D domain for medical image registration by integrating global attention and cross-modal feature fusion. In the field of medical image segmentation, several RWKV-based approaches have also been explored. For instance, RWKV-Unet \cite{jiang2025rwkv} integrates multi-scale global information via an inverted residual RWKV encoder and introduces a Cross-Channel Mix mechanism within the skip connections. Zig-RiR \cite{chen2025zig} enhances the spatial continuity of feature representations by introducing a novel spatial mixing mechanism and the Zig-Scan module. Zhou et al. \cite{zhou2024bsbp} proposed BSBP-RWKV, which incorporates Perona–Malik diffusion to preserve boundary features while suppressing background noise. Ye et al. \cite{ye2025hfe} introduced HFE-RWKV, along with a space-frequency consistency loss, to achieve accurate and efficient segmentation of pediatric left ventricular ultrasound images.

Although VRWKV has made progress in the field of medical image segmentation, several key issues remain to be addressed. First, most existing methods adopt hybrid architectures that combine CNN and RWKV components, as illustrated in Fig. \ref{fig2} (g). Due to structural modifications to the original VRWKV encoder, these approaches cannot effectively transfer large-scale pretrained VRWKV weights and lack a thorough exploration of the performance of pure VRWKV architectures. Second, existing studies do not sufficiently and systematically exploit the importance differences among frequency-domain information and multi-scale features, which may be highly beneficial for segmentation tasks. In contrast, as shown in Fig. \ref{fig2} (h), our proposed Med-URWKV adopts a pure RWKV architecture that is fully compatible with pretrained VRWKV encoders, facilitating an in-depth investigation of the potential of large-scale pretrained pure VRWKV models for medical image segmentation. Furthermore, by leveraging the intrinsic attention mechanism of VRWKV, we integrate frequency information and multi-scale features into the feature learning process and construct Med-URWKV$\dagger$ to further enhance segmentation performance.

\section{Methods}

\subsection{Overview of VRWKV}
\textbf{Input preprocessing.} This stage aims to transform input 2D images into sequential representations compatible with the model architecture. Specifically, an input image $I \in \mathbb{R}^{H \times W \times 3}$ is partitioned into $T = \frac{H \times W}{P^2}$ non-overlapping patches, where $P$ denotes the patch size and $T$ signifies the resultant token count. Subsequently, these patches are mapped into a $C$-dimensional latent space via a linear projection, and the integration of learnable position embeddings yields the final sequence features $X \in \mathbb{R}^{T \times C}$.
\begin{figure}[pos=htbp]
\includegraphics[width=0.78\textwidth,trim=6cm 13cm 2.3cm 12.2cm,clip]{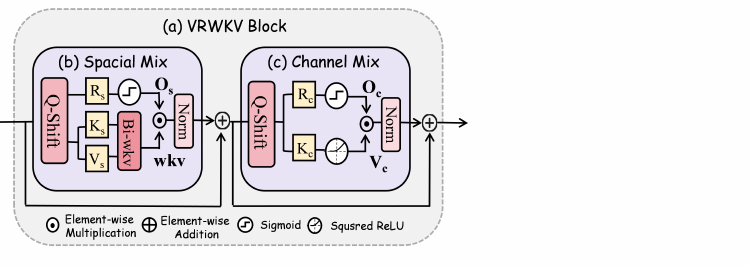}
\caption{(a) The internal structure of VRWKV Block. (b) The process of Spatial Mix. (c) The process of Channel Mix. } 
\label{fig3}
\end{figure}

\textbf{Spatial Mix module and Channel Mix module.} The schematic of the VRWKV block \cite{duan2024vision} is depicted in Fig. \ref{fig3} (a). Q-Shift serves as a token interaction mechanism tailored for vision-centric tasks, which effectively broadens the receptive field of individual tokens and incorporates local inductive biases for subsequent attention mechanisms. Specifically, it executes four-directional spatial shifts on input tokens $X$, concatenating feature slices from adjacent tokens with the original representation:

\begin{equation}
\begin{aligned}
\text{Q-Shift}_{(*)} (X) &= X + \left(1 - \mu_{(*)}\right) X^{\dagger}
\end{aligned}
\end{equation}
Here, $X^{\dagger}$ denotes the shifted features, which are obtained through slicing operations. $\ast \in \{R, K, V\}$ corresponds to the three matrices used in subsequent computations respectively, and $\mu_{(*)}$ is a learnable vector that controls the fusion ratio between the original and the shifted features.

As illustrated in Fig. \ref{fig3} (b), tokens are first fed into the Spatial Mix module to capture global dependencies, which is centered on the Bi-WKV mechanism. Specifically, the tokens processed by Q-Shift are projected through three parallel linear layers to generate matrices $R_s, K_s, V_s \in \mathbb{R}^{T \times C}$ as formulated in Equation \ref{eq1}, where $W_R, W_K$, and $W_V$ denote the corresponding learnable weight matrices.

\begin{equation}
\begin{gathered} 
R_s = \text{Q-shift}_R(X)W_R, \\ 
\begin{aligned} 
K_s = \text{Q-shift}_K(X)W_K, \quad V_s = \text{Q-shift}_V(X)W_V 
\end{aligned}
\end{gathered}
\label{eq1}
\end{equation}

Then $K_s$ and $V_s$ are used to compute the attention operator $wkv$ through the bi-WKV mechanism: 

\begin{equation}
\begin{split} 
wkv &= \frac{\sum_{i=0, i \neq t}^{T-1} e^{-(|t-i|-1)/T \cdot w + k_i} v_i + e^{u + k_t} v_t}{\sum_{i=0, i \neq t}^{T-1} e^{-(|t-i|-1)/T \cdot w + k_i} + e^{u + k_t}}.
\end{split}
\label{eq4}
\end{equation}

As formulated in Equation \ref{eq4}, the output for the $t$-th token is computed as a weighted sum of values across all tokens. These attention weights are jointly determined by the relative positional distance $(|t-i|-1)/T$, the decay vector $w$, and the key features $k_i$. Specifically, $u$ serves as a gain vector dedicated to the current token, ensuring the faithful preservation of its inherent features. Subsequently, $R_s$, activated via a Sigmoid function, functions as a gating mechanism to adaptively modulate the output response. The resulting features are then projected through a linear layer $W_O$ and stabilized by Layer Normalization (LN) to yield the final output representation:

\begin{equation}
O_s = (\sigma(R_s) \odot wkv) W_O
\label{eq3}
\end{equation}

Subsequently, the tokens are fed into the Channel Mix module. As illustrated in Fig. \ref{fig3} (c), this module functions as a Feed-Forward Network (FFN) designed to facilitate feature interactions across different channels. Following the identical Q-Shift processing procedure, the tokens are projected into $R_c$ and $K_c$ through two separate linear layers, as formulated in Equation \ref{eq5}.

\begin{equation}
\begin{aligned}
&R_c = \text{Q-shift}_R(X)W_R\\ 
&K_c = \text{Q-shift}_K(X)W_K
\end{aligned}
\label{eq5}
\end{equation}

The SquaredReLU activation function is applied to $K_c$ to enhance nonlinearity. $K_c$ is then passed through the linear layer $W_V$ to generate $V_C$, which is finally gated by $R_C$ and output via a linear layer: 

\begin{equation}
V_c = \text{SquaredReLU}(K_C)W_V
\label{eq7}
\end{equation}
\begin{equation}
O_c = (\sigma(R_c) \odot V_c) W_O
\label{eq6}
\end{equation}

Finally, the outputs of the above two modules are added to the inputs via residual connections to avoid gradient vanishing in deep networks.

\begin{figure*}[pos=htbp]
\centering
\includegraphics[width=\textwidth,trim=0cm 0cm 0cm 0cm,clip]{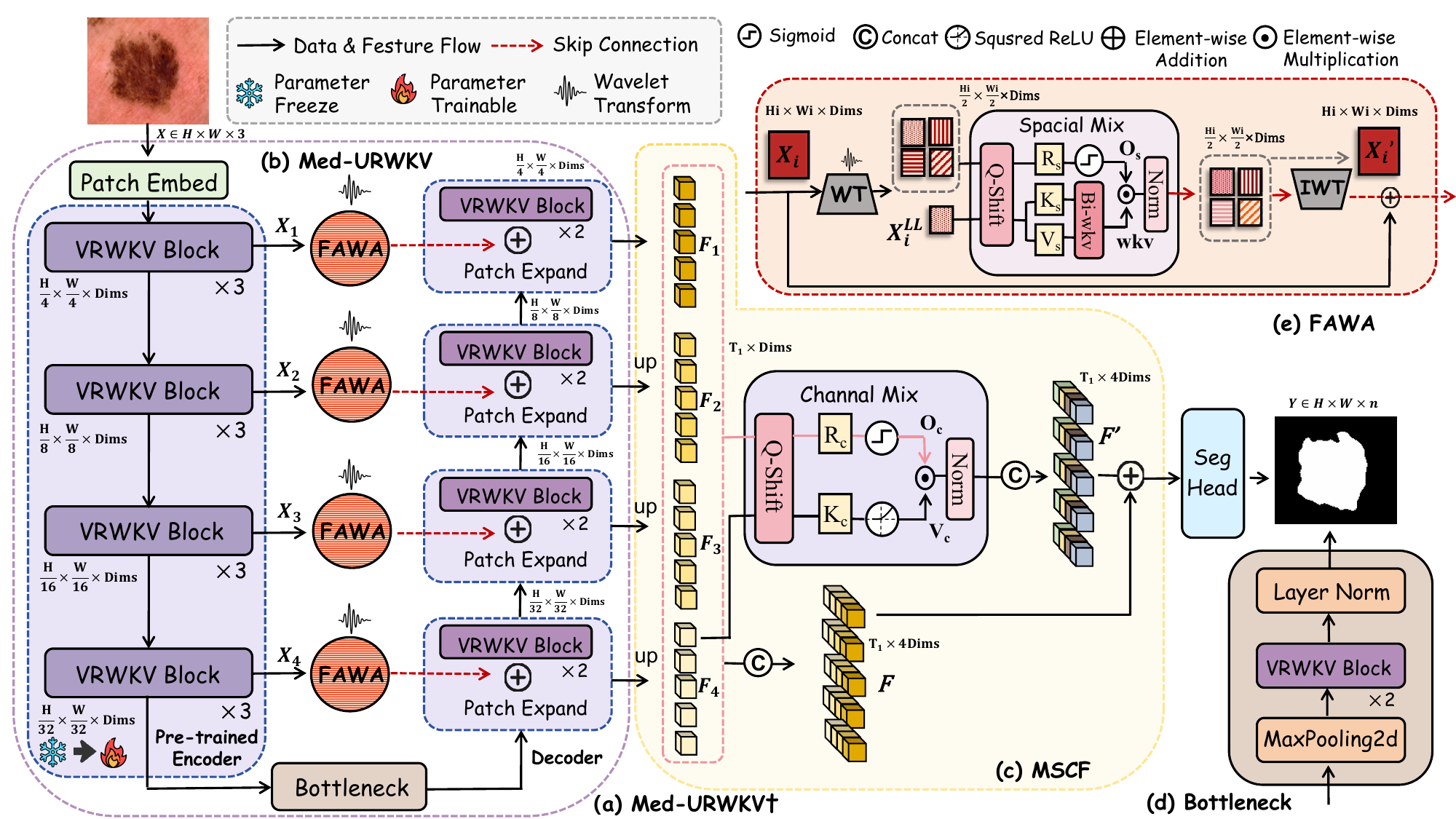}
\caption{Detailed components of the proposed framework: (a) Overview of the Med-URWKV$\dagger$ architecture. (b) The Med-URWKV, comprising a pretrained VRWKV encoder, a Bottleneck block, and a pure VRWKV decoder. (c) Structure of the Multi-Scale Channel Fusion (MSCF) module. (d) Detailed view of the Bottleneck block. (e) Structure of the Frequency-Aware Wavelet Attention (FAWA) module.} 
\label{fig4}
\end{figure*}

\subsection{Overall Architecture of Proposed Models}
\subsubsection{Med-URWKV}
As illustrated in Fig. \ref{fig4} (b), the proposed Med-URWKV architecture comprises three primary modules: an efficiently reusable pretrained VRWKV encoder, a corresponding decoder, and a VRWKV-based bottleneck bridging the two. Specifically, an input image $X\in \mathbb{R}^{H\times W \times C}$ is initially processed by the VRWKV encoder on the left of Fig. \ref{fig4} (b) to extract four hierarchical multi-scale features $X_i \in \mathbb{R}^{\frac{H}{2^{i+1}}\times \frac{W}{2^{i+1}} \times \text{Dims}}$ with $i=1,2,3,4$, where $H$ and $W$ denote the spatial dimensions and $\text{Dims}$ represents the predefined embedding dimensionality of the pretrained backbone. The output features $X_4$ from the final encoder stage are subsequently channeled through the bottleneck block, as shown in Fig. \ref{fig4} (d), for further feature abstraction and dimensionality reduction. These refined features serve as the input to the decoder, where progressive patch expansion and feature decoding are performed. Meanwhile, the hierarchical encoder features are integrated via skip connections to preserve spatial details. Finally, the top-layer decoder output is processed by a segmentation head utilizing a $1\times1$ convolutional layer to calibrate the channel dimensions, ultimately generating the segmentation map $Y \in\mathbb{N}^{H\times W \times n}$, where $n$ denotes the number of semantic classes.

\subsubsection{Med-URWKV$\dagger$}
The complete architecture of Med-URWKV$\dagger$ is illustrated in Fig. \ref{fig4}(a). Based on Med-URWKV, the hierarchical features $X_i$ are individually processed by the FAWA modules to generate feature representations with enhanced structural stability and richer detail information. Both the bottleneck layer and the decoder remain unchanged, with the decoder outputs still represented in the form of tokens, denoted as $F_i \in \mathbb{R}^{T_i \times \text{Dims}}$. Subsequently, these four are fed into the MSCF module for multi-scale channel interaction, and output $F_{out}$. Finally, the output features of the MSCF are first upsampled to restore the original height and width of the feature map, and then fed into the segmentation head for subsequent processing.
%
The segmentation result $Y \in \mathbb{N}^{H\times W \times n}$ with the complete calculation process are presented in Equation \ref{eq14}.
\begin{equation}
Y=\text{Seg Head}(\text{Patch Expand} (F_{out})))
\label{eq14}
\end{equation}

\vspace{-0.05cm}
\begin{figure}[pos=htbp]
\includegraphics[width=\columnwidth,trim=6.2cm 7cm 13cm 5.5cm,clip]{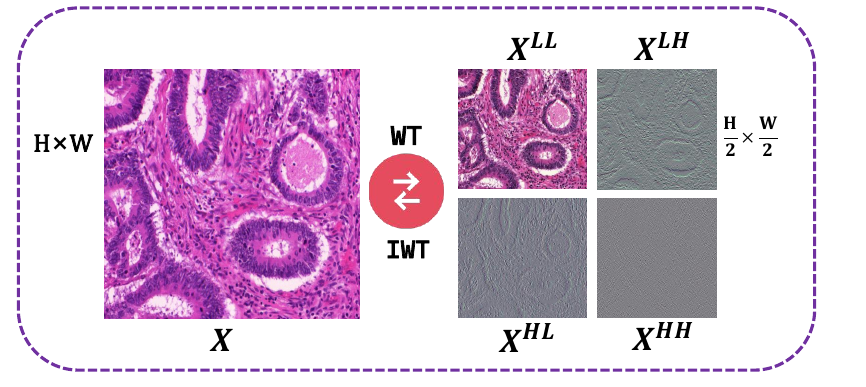}
\caption{The process of first-order wavelet transform and inverse wavelet transform. Image taken from GLAS \cite{sirinukunwattana2017gland}} \label{fig5}
\vspace{-0.05cm}
\end{figure}

\subsection{Frequency-Aware Wavelet Attention (FAWA) Module}
As shown in Fig. \ref{fig5}, Wavelet Transform (WT) \cite{ghazali2007feature} decomposes images into multi-scale sub-bands: low-frequency components preserve global contours, while high-frequency components capture detailed information such as edges and textures. Compared with traditional methods such as the Fourier transform, WT can more efficiently generate low-frequency and high-frequency signals \cite{zhou2023xnet}, it can localize image features in both the spatial and frequency domains simultaneously. Inspired by the concept of the Frequency Attention Triplet \cite{huang2025wavelet}, we note that the WKV attention mechanism inherent to VRWKV can also be applied in the frequency domain. Therefore, we propose the Frequency-Aware Wavelet Attention (FAWA) module, whose architecture is illustrated in Fig. \ref{fig4} (e). In FAWA, We select the low-frequency sub-band as the key and value in the attention mechanism because it preserves global structural and information with high stability, which is critical for modeling long-range dependencies in medical images. In contrast, high-frequency sub-bands mainly capture local variations and edge cues and are therefore suitable for generating attention weights. This design enables high-frequency details to be selectively enhanced under low-frequency guidance, thereby improving boundary precision.

First, it receives $X_i$ from defferent layers of the encoder and performs haar WT on it. Each $X_i$ generates four frequency sub-band feature maps ${X_i}^{LL}, {X_i}^{LH}, {X_i}^{HL}, {X_i}^{HH}$, which correspond to the low-frequency component, vertical, horizontal and horizontal high-frequency component, respectively. The spatial dimensions of each sub-band feature map are reduced to half of the original size through the transformation shown in Equation \ref{eq9}.
\begin{equation}
\begin{aligned}
{X_i}^{LL}, {X_i}^{LH}, {X_i}^{HL}, {X_i}^{HH} &= \mathrm{WT}(X_i), \quad i=1,2,3,4 
\end{aligned}
\label{eq9}
\end{equation}

Subsequently, the four frequency sub-band feature maps ${X_i}^{LL}, {X_i}^{LH}, {X_i}^{HL}, {X_i}^{HH} \in \mathbb{R}^{\frac{H_i}{2} \times \frac{W_i}{2} \times \text{Dims}}$ are fed as the receive vector $R_s$, while the low-frequency component ${X_i}^{LL}$ is simultaneously used as the key vector 
$K_s$ and value vector $V_s$ and input into the Spatial Mix module. The RWKV attention mechanism is adopted to finely adjust the multi-frequency features, $R_s$ converts frequency information into a weight distribution, and then uses this distribution to filter low-frequency values. 
\begin{equation}
\begin{aligned}
{{X_i}^j}' &= \mathrm{SpatialMix}({X_i}^j,{X_i}^{LL}),\\ &j=LL,LH,HL,HH
\end{aligned}
\label{eq10}
\end{equation}
As illustrated in Equation \ref{eq10}, the resulting values match the global structure and preserve high-frequency characteristics, thus enabling high-frequency components to reference low-frequency global structural information and measure the importance of different frequency features to overall spatial features.

Finally, the feature maps ${{X_i}^{LL}}', {{X_i}^{LH}}', {{X_i}^{HL}}', {{X_i}^{HH}}’$ after attention adjustment are restored to the size of the original feature map via inverse wavelet transform, generating a fused feature map $X'_i$. The final output of the FAWA module is $X'_i+X_i$, which preserves the basic information of the original features and enhances the stability of training. The relevant calculation formula is shown in Equation \ref{eq11}. 
\begin{equation}
\begin{aligned}
X'_i &= \mathrm{IWT}({{X_i}^{LL}}', {{X_i}^{LH}}', {{X_i}^{HL}}', {{X_i}^{HH}}’), \quad i=1,2,3,4 \\
& X_{out}=X'_i+X_i
\end{aligned}
\label{eq11}
\end{equation}
\subsection{Multi-Scale Channel Fusion (MSCF) Module}
In medical imaging, the morphological scale of lesions exhibits significant intra-class variability, posing challenges for comprehensive representation via single-scale features. Furthermore, medical images often contain substantial spatial redundancy, and since lesion regions are typically highly localized, the direct utilization of global features may lead to the attenuation of salient information. To address these issues, we propose the MSCF module, which adaptively learns instance-specific scale weights for each input image. This mechanism enables the network to prioritize the most effective scale features for the current segmentation task while simultaneously enhancing discriminative channel representations.

In Fig. \ref{fig4} (c), We define the outputs of the four decoding layers as $F_1 \in \mathbb{R}^{T_1 \times \text{Dims}}$, $F_2 \in \mathbb{R}^{T_2 \times \text{Dims}}$, $F_3 \in \mathbb{R}^{T_3 \times \text{Dims}}$, and $F_4 \in \mathbb{R}^{T_4 \times \text{Dims}}$. 
MSCF module first performs upsampling on $F_2-F_4$ to the same size as $F_1$, which lays the foundation for subsequent cross-scale fusion.

Based on scale alignment, the module introduces the Channel Mix mechanism from the VRWKV. Taking $F_4$ as the reference anchor, channel-dimension interaction is performed on the aligned $F_1$, $F_2$, $F_3$, $F_4$, where $F_4$ acts as $K_c$ input and $F_1- F_4$ serve as $R_c$ inputs. $F_4$ possesses the smallest spatial resolution and carries the original global semantic information, then $F1-F4$ rely on their own information to perform channel-wise weight assignment and selective activation on the global content refined by $F_4$, this enables the features at each level to have gating effects. The final output can reflect the relative importance of each channel feature for the task at the current scale.

Ultimately, a residual fusion strategy is adopted to further optimize feature representation, the detailed calculation process is shown in Equation \ref{eq12}. The scale-aligned original feature maps are first concatenated along the channel dimension to form a base feature 
$F\in \mathbb{R}^{T_1 \times \text{4Dims}}$ that preserves complete multi-scale information. Meanwhile, the cross-scale interaction–enhanced feature maps are concatenated to obtain an enhanced feature $F'\in \mathbb{R}^{T_1 \times \text{4Dims}}$. The base and enhanced features are then combined by element-wise addition, preventing the loss of critical information during feature fusion. The final output $F_{out}$ is sent to the segmentation head to produce the final segmentation output $Y \in \mathbb{N}^{H\times W \times n}$.

\begin{equation}
\begin{aligned}
&F= \text{Concat} (F_i), \quad i=1,2,3,4\\
&F'= \text{Concat} (\text{ChannelMix} (F_i,F_4))\\
&F_{out}=F'+F
\end{aligned}
\label{eq12}
\end{equation}

\section{Experiments}
\subsection{Datasets}
We comprehensively evaluated Med-URWKV and Med-URWKV$\dagger$ across five publicly available benchmarks, encompassing: the ISIC2017 and ISIC2018 skin lesion segmentation datasets \cite{codella2018skin} ($2,000/150/600$ and $2,594/100/1,0\\00$ images for training, validation, and testing, respectively); the GLAS histopathology dataset ($80$ training and $85$ test images) \cite{sirinukunwattana2017gland}; the BUSI breast ultrasound dataset ($647$ images) \cite{al2020dataset}; and the KvasirSEG polyp segmentation dataset ($1,000$ RGB images) \cite{jha2019kvasir}. For the ISIC series and GLAS benchmarks, the official dataset partitioning schemes were strictly followed. The remaining datasets were randomly partitioned into training and test sets with an $8:2$ ratio, consistent with the experimental protocols established in prior work \cite{he2024frcnet}.

\begin{table*}[t]
\caption{Quantitative comparison of segmentation performance between the proposed Med-URWKV family and SOTA methods across five datasets. Bold red and underlined blue fonts indicate the best and second-best results, respectively. The $\downarrow$ signifies that lower values correspond to superior performance. To ensure statistical reliability, all experiments were conducted in triplicate, and the mean results are reported.}
\centering
\resizebox{\linewidth}{!}{
\begin{tabular}{c|c|c|cc|cc|cc|cc|cc|cc|c}
\specialrule{1pt}{0pt}{0pt}
\specialrule{1pt}{0pt}{0pt}
\multirow{2}{*}{\textbf{Type}} & \multirow{2}{*}{\textbf{Methods}} & \multirow{2}{*}{\textbf{Source}} & \multicolumn{2}{c}{\textbf{BUSI}}\vline & \multicolumn{2}{c}{\textbf{ISIC-2017}}\vline & \multicolumn{2}{c}{\textbf{ISIC-2018}}\vline & \multicolumn{2}{c}{\textbf{Kvasir-SEG}}\vline & \multicolumn{2}{c}{\textbf{GLAS}}\vline & \multicolumn{2}{c}{\textbf{Average Result}}\vline & \multirow{2}{*}{\textbf{Params. $\downarrow$}}\\
&&&DSC&IoU&DSC&IoU&DSC&IoU&DSC&IoU&DSC&IoU&DSC&IoU&\\
\specialrule{0.8pt}{1pt}{1pt}

\multirow{7}{*}{CNN} 
&UNet\cite{ronneberger2015u}&MICCAI'15&68.20&58.71&80.82&71.66&86.81&79.06&86.44&79.20&83.28&72.28&81.11&72.18&24.89 M\\
&Attention-UNet\cite{oktay2018attention}&MIDL'18&65.42&55.78&81.42&72.51&85.87&77.77&85.75&78.06&90.78&83.78&81.85&73.58&34.88 M\\
&UNet++\cite{zhou2019unet++}&TMI'19&66.41&56.28&80.90&71.72&84.26&75.94&85.77&77.96&90.13&82.95&81.49&72.97&47.19 M\\
&UNeXt\cite{valanarasu2022unext}&MICCAI'22&73.25&64.04&84.27&75.65&88.37&81.31&82.47&74.06&88.56&80.61&83.38&75.13&1.47 M\\
&MGFuseSeg\cite{xu2023mgfuseseg}&BIBM'23&69.36&60.05&82.79&74.11&88.97&82.00&80.37&71.97&87.41&78.97&81.78&73.42&22.27 M\\
&ACC-Unet\cite{ibtehaz2023acc}&MICCAI'23&76.43&67.75&82.35&73.81&87.25&79.84&85.87&79.12&87.92&79.63&83.96&76.03&16.77 M\\
&PVT-EMCAD-b1\cite{rahman2024emcad}&CVPR'24&78.43&69.62&83.56&74.75&88.59&81.59&86.49&79.35&85.71&76.01&84.56&76.26&15.41 M\\

\specialrule{0.8pt}{1pt}{1pt}

\multirow{4}{*}{ViT}
&TransUNet\cite{chen2021transunet}&Arxiv'21&76.44&67.04&84.44&75.87&89.38&82.56&85.46&78.32&85.29&75.95&84.20&75.95&92.23 M\\
&Swin-Unet\cite{cao2022swin}&ECCV'22&49.58&38.48&83.27&74.20&77.44&67.51&56.35&43.37&81.20&69.48&69.57&58.61&27.15 M\\
&UCTransNet\cite{wang2022uctransnet}&AAAI'22&75.51&66.34&82.08&73.55&87.80&80.57&86.42&79.17&89.64&82.30&84.29&76.39&66.24 M\\
&MISSFormer\cite{huang2021missformer}&TMI'23&74.16&63.87&81.32&72.54&88.30&81.09&71.81&61.62&84.24&74.14&79.97&70.65&35.45 M\\

\specialrule{0.8pt}{1pt}{1pt}

\multirow{3}{*}{Mamba}
&VM-UNet\cite{ruan2024vm}&Arxiv'24&61.35&50.13&83.68&75.19&89.20&82.24&70.12&61.10&89.80&82.50&78.83&70.23&34.62 M\\
&Swin-UMamba\cite{liu2024swin}&MICCAI'24&\textcolor{red}{\textbf{82.24}}&\textcolor{red}{\textbf{73.34}}&84.86&76.38&88.97&81.92&84.03&76.32&90.91&84.28&86.20&78.45&55.06 M\\
&H-Vmnet\cite{wu2403h}&Neurocomputing'25&74.44&64.02&84.48&76.01&88.75&81.61&75.99&66.01&82.28&71.51&81.18&71.83&6.44 M\\

\specialrule{0.8pt}{1pt}{1pt}

\multirow{6}{*}{RWKV}
&Zig-RiR\cite{chen2025zig}&TMI'25&73.87&63.91&84.55&75.92&88.79&82.11&86.31&75.21&89.04&81.13&84.51&75.66&24.58 M\\
&HFE-RWKV\cite{ye2025hfe}&ICASSP'25&53.14&41.36&79.17&69.30&85.25&77.24&50.34&37.67&71.13&56.40&67.81&56.39&49.67 M\\
&RWKV-UNet\cite{jiang2025rwkv}&Arxiv'25&77.13&67.30&84.89&76.26&89.18&82.17&85.56&78.06&90.57&83.53&85.47&77.86&17.40 M\\
\cline{2-16}
&Med-URWKV-T&Ours&77.56&67.84&\textcolor{blue}{\underline{84.91}}&\textcolor{blue}{\underline{76.44}}&\textcolor{red}{\textbf{89.85}}&\textcolor{blue}{\underline{82.91}}&\textcolor{blue}{\underline{90.35}}&\textcolor{blue}{\underline{84.46}}&91.02&84.24&\textcolor{blue}{\underline{86.74}}&79.18&14.33 M\\
&Med-URWKV-S&Ours&78.46&69.63&84.51&76.10&89.53&82.66&89.89&84.11&\textcolor{blue}{\underline{91.13}}&\textcolor{blue}{\underline{84.42}}&86.70&\textcolor{blue}{\underline{79.38}}&53.07 M\\
&Med-URWKV$\dagger$&Ours&\textcolor{blue}{\underline{80.90}}&\textcolor{blue}{\underline{72.42}}&\textcolor{red}{\textbf{85.91}}&\textcolor{red}{\textbf{77.69}}&\textcolor{blue}{\underline{89.80}}&\textcolor{red}{\textbf{82.95}}&\textcolor{red}{\textbf{91.41}}&\textcolor{red}{\textbf{85.96}}&\textcolor{red}{\textbf{91.97}}&\textcolor{red}{\textbf{85.76}}&\textcolor{red}{\textbf{88.00}}&\textcolor{red}{\textbf{80.96}}&24.41 M\\
\specialrule{1pt}{0pt}{0pt}
\specialrule{1pt}{0pt}{0pt}
\end{tabular}
}
\label{tab1}
\end{table*}

\subsection{Implementation Details}
We evaluated Med-URWKV at two varying scales, namely Med-URWKV-T and Med-URWKV-S, incorporating their respective pretrained VRWKV encoders. Specifically, the Tiny and Small variants utilize the VRWKV official pre-trained weight \texttt{upernet\_vrwkv\_adapter\_tiny\_512\_\\160k\_ade20k} and \texttt{upernet\_vrwkv\_adapter\_small\_512\_160k\_ade\\20k} backbones, respectively. Only the encoder backbones and their corresponding weights were preserved, while irrelevant components from the pretraining phase were discarded. The remaining modules of our architecture were initialized using a standard random initialization scheme. Regarding competitive baselines, PVT-EMCAD-b1, Swin-UMamba, and RWKV-UNet were initialized with their official pretrained weights, while all other comparative methods were trained from scratch with random initialization.

During training on all datasets, the pretrained encoders of Med-URWKV and Med-URWKV$\dagger$ were frozen for the first 10 epochs to facilitate weight alignment. Afterward, the encoder parameters were unfrozen and jointly optimized with the rest of the network. All input images were resized to a resolution of $[512, 512]$ and processed in RGB format. Data augmentation included random horizontal flipping, random vertical flipping, and random rotation. The AdamW optimizer was employed with an initial learning rate of $3\times10^{-4}$ and a weight decay of $1\times10^{-4}$. The batch size was set to 8, and the maximum number of training epochs was 200. An early stopping strategy was adopted to prevent overfitting: training was terminated if the performance on the test set did not improve for 10 consecutive epochs. All experiments on each dataset were repeated three times, and the average results were reported. The training loss was defined as a combination of cross-entropy loss and Dice loss, while Intersection over Union (IoU) and Dice Similarity Coefficient (DSC) were used as evaluation metrics. To ensure fairness, all experiments were conducted under identical experimental settings. The models were implemented using the PyTorch framework and trained on two NVIDIA RTX A6000 GPUs with 48 GB of memory each.

\subsection{Comparison with Other Methods}
As shown in Table. \ref{tab1}, the Med-URWKV constructed from a pretrained pure VRWKV backbone demonstrates strong transferability in medical image segmentation tasks, achieving second-best or even best performance on most datasets and evaluation metrics. This clearly indicates that, compared with carefully designed hybrid RWKV architectures trained from scratch, pretrained pure RWKV models exhibit greater potential for medical image segmentation, further validating the feasibility of applying VRWKV to this domain. Building upon this foundation, the optimized Med-URWKV$\dagger$ achieves further performance gains, delivering the best results on four out of the five datasets. In terms of overall average performance, Med-URWKV† and Med-URWKV-S rank first and second, achieving average DSC scores of 88.00\% and 86.74\%, respectively, and outperforming the remaining hybrid RWKV-based methods. Furthermore, Med-URWKV† improves the average DSC by 1.80\% compared with Swin-UMamba, which exhibits the best performance among the alternative architectures, while using less than half of its parameters. These results indicate the strong potential of VRWKV architectures for medical image segmentation and demonstrate that the proposed improvements achieve favorable performance gains with high parameter efficiency.

As shown in Table. \ref{tab1} and Fig. \ref{fig1}, Med-URWKV models at different scales achieve comparable performance. This suggests that although pure VRWKV exhibits considerable potential in medical image segmentation, simply increasing the number of parameters is no longer an optimal strategy for achieving efficient and high-performance deployment. In contrast, the improved Med-URWKV$\dagger$ achieves superior segmentation performance using less than half the parameters of Med-URWKV-S, clearly highlighting the effectiveness of FAWA and MSCF in efficiently elevating the performance ceiling of pure VRWKV models in medical imaging applications.

\begin{figure*}[pos=htbp]
\includegraphics[width=\textwidth,trim=0cm 4cm 0cm 2cm,clip]{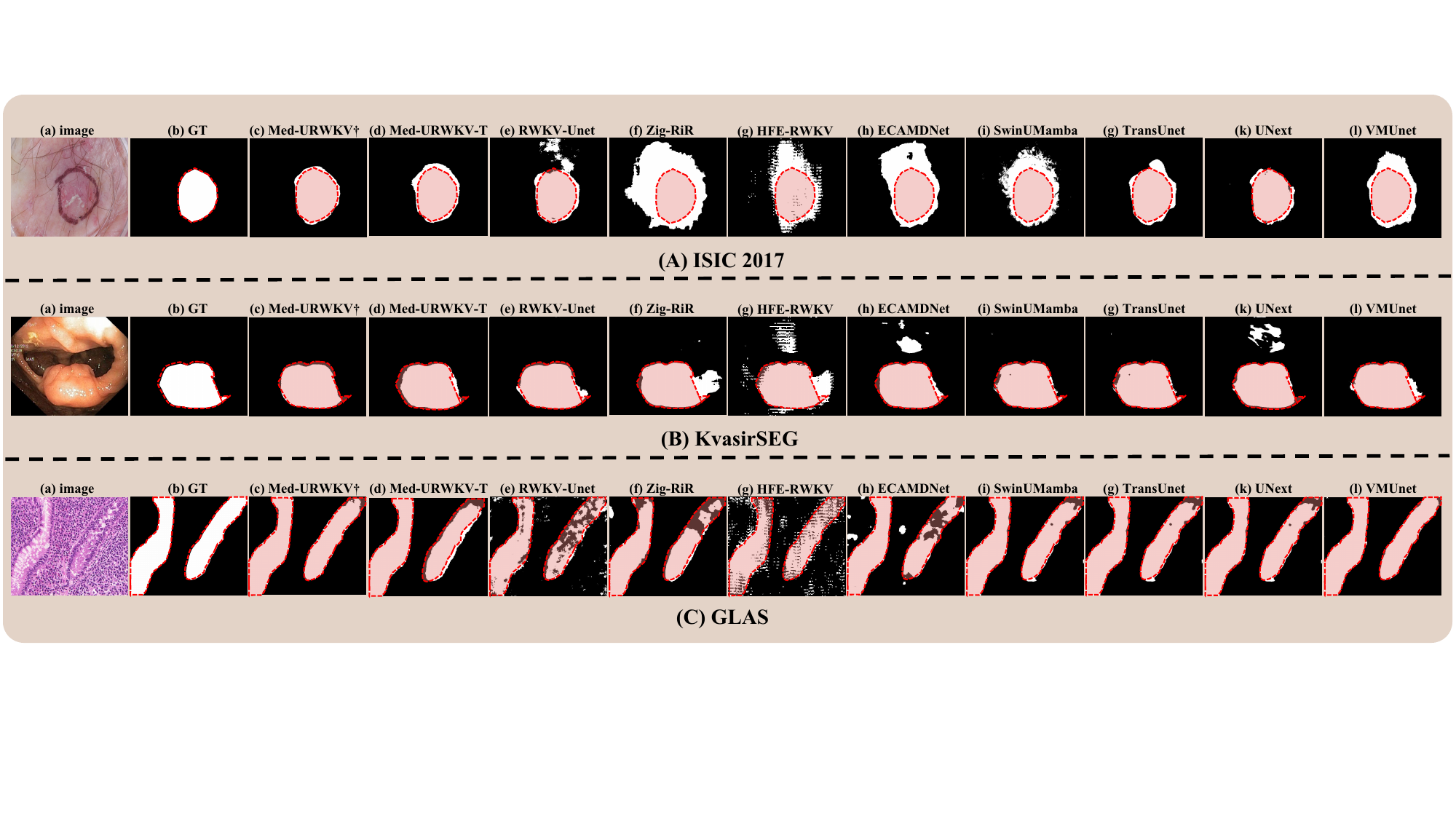}
\caption{Qualitative visual comparison of Med-URWKV-T and Med-URWKV$\dagger$ with other competing models. GT denotes the ground truth. White pixels inside the light red region and black pixels outside the region indicate correct predictions, while all other cases represent incorrect predictions.} \label{fig6}
\end{figure*}

Fig. \ref{fig6} provides a qualitative visualization comparing the segmentation masks generated by Med-URWKV$\dagger$ and Med-URWKV-T against representative state-of-the-art methods across three datasets. In Fig. \ref{fig6} (A), skin lesion segmentation is hindered by indistinct boundaries and high-frequency noise derived from hair artifacts \cite{zhou2024spatial}. Med-URWKV$\dagger$ exhibits superior robustness to these disturbances, yielding segmentation contours that demonstrate high fidelity to the ground truth while remaining largely unaffected by hair-induced noise over the lesion regions. In the polyp segmentation task shown in Fig. \ref{fig6} (B), where boundaries are often ambiguous and prone to confusion with surrounding colon tissues, Med-URWKV$\dagger$ effectively discriminates polyp regions from the background, achieving results with the highest consistency to the ground truth. Fig. \ref{fig6} (C) illustrates performance on the GLAS dataset for cellular recognition. Due to the diminished contrast in microscopic imaging and interference from clustered blood cells, cellular boundaries are particularly arduous to delineate. Despite these challenges, Med-URWKV$\dagger$ maintains robust performance. Conversely, other methods, such as HFE-RWKV \cite{ye2025hfe} and Zig-RiR \cite{chen2025zig}, suffer from varying degrees of interference, whereas Med-URWKV$\dagger$ effectively mitigates such artifacts and produces predictions that closely align with the ground truth even under severe low-contrast conditions.

\subsection{Ablation Studies}
To systematically validate the efficacy of the proposed pretraining strategy, as well as the FAWA and MSCF modules, we perform a series of ablation studies and provide a comprehensive analysis to gain deeper insights into the individual contribution of each architectural component to the overall performance.

\begin{table}[pos=htbp]
\caption{The ablation study on the pretrained VRWKV encoder. w/ denotes the use of a VRWKV encoder initialized with pretrained weights, w/o refers to a VRWKV encoder without pretrained initialization, where the parameters are randomly initialized and trained from scratch. The experiments were conducted on three datasets: BUSI, Kvasir-SEG, and GLAS, using DSC as the evaluation metric.}

\centering
\begin{tabular}{c|c|c|c}
\specialrule{1pt}{0pt}{0pt}
\specialrule{1pt}{0pt}{0pt}
Dataset&Epoch&w/ pre-training&w/o pre-training\\
\hline
\multirow{4}{*}{BUSI}&10&31.04&19.47 \\
&50&75.77&39.85 \\
&100&77.34&49.19 \\
\cline{2-4}
&\textbf{best}&\textbf{77.56}&\textbf{51.17} \\
\specialrule{1pt}{0pt}{0pt}
\multirow{4}{*}{Kvasir-SEG}&10&55.93&46.99 \\
&50&87.97&59.03 \\
&100&90.15&66.29 \\
\cline{2-4}
&\textbf{best}&\textbf{90.35}&\textbf{69.63} \\
\specialrule{1pt}{0pt}{0pt}
\multirow{4}{*}{GLAS}&10&52.04&48.59 \\
&50&85.89&64.37 \\
&100&90.55&66.87 \\
\cline{2-4}
&\textbf{best}&\textbf{91.02}&\textbf{68.83} \\
\specialrule{1pt}{0pt}{0pt}
\specialrule{1pt}{0pt}{0pt}
\end{tabular}
\label{tab2}
\end{table}
\subsubsection{Pre-training VRWKV encoder}
We performed a series of experiments using Med-URWKV-T to evaluate the efficacy of the pretrained VRWKV encoder on the BUSI, Kvasir-SEG, and GLAS datasets, with the corresponding quantitative results summarized in Table \ref{tab2}. The experimental results clearly demonstrate that Med-URWKV-T, when integrated with the pretrained VRWKV encoder, consistently surpasses its randomly initialized counterpart across all metrics, yielding a substantial performance gain of up to 26.39\%. This underscores that the pretrained VRWKV encoder is instrumental in enhancing performance for complex medical image segmentation tasks. Notably, during the initial ten epochs when the encoder parameters remain frozen—restricting training to the bottleneck and decoder modules—the architecture with the pretrained encoder already exhibits superior segmentation capability. Upon unfreezing the encoder at subsequent stages, the pretrained backbone consistently facilitates accelerated convergence and achieves higher peak segmentation accuracy across all three benchmarks. These findings further confirm the pivotal role of utilizing a pretrained VRWKV encoder for robust feature extraction.

\subsubsection{Analysis of FAWA and MSCF}
To validate the effectiveness of the proposed FAWA and MSCF modules in enhancing the performance of the pure VRWKV baseline, we adopt Med-URWKV-T as the baseline model and employ a pretrained encoder to conduct ablation studies on the BUSI and Kvasir-SEG datasets. The training and validation strategies are kept consistent with those used in the comparative experiments. DSC and IoU are used as evaluation metrics, with results summarized in Table \ref{tab3}.
\begin{table}[pos=htbp]
  \centering
  \caption{Quantitative ablation analysis evaluating the individual and joint efficacy of the proposed FAWA and MSCF modules. The symbols "$-$" and "$\checkmark$" denote the exclusion and inclusion of the respective modules. Experiments were performed on the BUSI and Kvasir-SEG benchmarks, with DSC and IoU employed as the evaluation metrics.}
  \label{tab_method_perf}
  \setlength{\tabcolsep}{2.7pt}
  \begin{tabular}{c|c|c|cc|cc}  
    \specialrule{1pt}{0pt}{0pt} 
    \multicolumn{3}{c|}{Method} & \multicolumn{2}{c|}{BUSI} & \multicolumn{2}{c}{KvasirSEG} \\ 
    \cline{1-7} 
    Med-URWKV-T & FAWA & MSCF & DSC & IOU & DSC & IOU \\  
    \specialrule{0.5pt}{0pt}{0pt} 
     \checkmark & - & - & 77.56 & 67.84 & 90.35 & 84.46 \\ 
     \checkmark & \checkmark & - & 78.66 & 69.33 & 90.69 & 85.09 \\
     \checkmark & - & \checkmark & 80.30 & 71.56 & 90.75 & 85.38 \\
     \checkmark & \checkmark & \checkmark & \textbf{80.90} & \textbf{72.42} & \textbf{91.41} & \textbf{85.96} \\
    \specialrule{1pt}{0pt}{0pt}  
  \end{tabular}
  \label{tab3}
\end{table}

Four experimental settings are designed to investigate the individual contributions of each module as well as their combined effect. By comparing the first and second rows of Table \ref{tab3}, it can be observed that introducing the FAWA module into the skip connections of Med-URWKV-T leads to improvements of 1.10\% and 1.49\% in DSC and IoU on the BUSI dataset, respectively, and gains of 0.34\% and 0.63\% on the Kvasir-SEG dataset. These results demonstrate the effectiveness of FAWA, which decomposes feature representations in the frequency domain via wavelet transform and leverages the VRWKV attention mechanism for adaptive modulation, enabling better alignment between high-frequency details and low-frequency global structures, thereby improving segmentation accuracy. As shown in the third row, when only the MSCF module is employed, segmentation performance improves on both datasets, with more pronounced gains observed in scenarios characterized by large variations in lesion size and prominent local features, such as the BUSI dataset. When FAWA and MSCF are jointly applied (the last row of Table \ref{tab3}), the resulting model, termed Med-URWKV$\dagger$, achieves the best performance across both datasets. These results indicate that the two modules are complementary and jointly indispensable: FAWA primarily alleviates the mismatch between local details and global structural information, while MSCF enhances effective multi-scale feature fusion. Their combination ultimately leads to progressive and consistent performance improvements.

\section{Conclusion and Future Work}
This work systematically investigates the potential of large-scale pretrained pure VRWKV models for medical image segmentation. Two variants, Med-URWKV-T and Med-URWKV-S, are constructed to evaluate the effect of model scale. Experimental results demonstrate that pure VRWKV architectures achieve performance comparable to or exceeding SOTA methods across multiple benchmarks. However, we observe that simply increasing model parameters leads to diminishing gains in segmentation accuracy under clinical settings. Based on these findings, we propose Med-URWKV$\dagger$, an enhanced pure VRWKV model that integrates two key modules, FAWA and MSCF. These modules enable frequency-aware global dependency modeling and adaptive multi-scale feature fusion. Extensive evaluations on five public datasets against 17 competing methods show that Med-URWKV$\dagger$ consistently outperforms mainstream approaches. Notably, while retaining a pure VRWKV backbone, Med-URWKV$\dagger$ surpasses several hybrid RWKV-based models and achieves superior segmentation performance compared to Med-URWKV-S with less than half the parameters. Ablation studies further confirm the critical role of the pretrained VRWKV encoder and the complementary effects of FAWA and MSCF.

Future work will explore the applicability of pure VRWKV architectures to broader medical imaging tasks, imaging modalities, and clinical domains, as well as their extension to classification and reconstruction tasks. In addition, advanced learning paradigms such as semi-supervised and self-supervised learning will be investigated to further enhance generalization and clinical robustness.

\section*{Acknowledgments}
This work is partially supported by the National Natural Science Foundation (62272248), the Natural Science Foundation of Tianjin (23JCZDJC01010).

\section*{Declaration of generative AI and AI-assisted technologies in the manuscript preparation process}
During the preparation of this work the authors used ChatGPT in order to complete English translation, polishing and latex typesetting assistance. After using this tool, the authors reviewed and edited the content as needed and take full responsibility for the content of the published article.

\bibliographystyle{cas-model2-names}
\bibliography{ref}

@inproceedings{ronneberger2015u,
  title={U-net: Convolutional networks for biomedical image segmentation},
  author={Ronneberger, Olaf and Fischer, Philipp and Brox, Thomas},
  booktitle={Medical image computing and computer-assisted intervention--MICCAI 2015: 18th international conference, Munich, Germany, October 5-9, 2015, proceedings, part III 18},
  pages={234--241},
  year={2015},
  organization={Springer}
}

@article{jiang2025rwkv,
  title={RWKV-UNet: Improving UNet with Long-Range Cooperation for Effective Medical Image Segmentation},
  author={Jiang, Juntao and Zhang, Jiangning and Liu, Weixuan and Gao, Muxuan and Hu, Xiaobin and Yan, Xiaoxiao and Huang, Feiyue and Liu, Yong},
  journal={arXiv preprint arXiv:2501.08458},
  year={2025}
}

@article{chen2025zig,
  title={Zig-RiR: Zigzag RWKV-in-RWKV for Efficient Medical Image Segmentation},
  author={Chen, Tianxiang and Zhou, Xudong and Tan, Zhentao and Wu, Yue and Wang, Ziyang and Ye, Zi and Gong, Tao and Chu, Qi and Yu, Nenghai and Lu, Le},
  journal={IEEE Transactions on Medical Imaging},
  year={2025},
  publisher={IEEE}
}

@inproceedings{zhou2024bsbp,
  title={Bsbp-rwkv: Background suppression with boundary preservation for efficient medical image segmentation},
  author={Zhou, Xudong and Chen, Tianxiang},
  booktitle={Proceedings of the 32nd ACM International Conference on Multimedia},
  pages={4938--4946},
  year={2024}
}

@inproceedings{ye2025hfe,
  title={HFE-RWKV: High-Frequency Enhanced RWKV Model for Efficient Left Ventricle Segmentation in Pediatric Echocardiograms},
  author={Ye, Zi and Chen, Tianxiang and Wang, Ziyang and Zhang, Hanwei and Zhang, Lijun},
  booktitle={ICASSP 2025-2025 IEEE International Conference on Acoustics, Speech and Signal Processing (ICASSP)},
  pages={1--5},
  year={2025},
  organization={IEEE}
}

@article{duan2024vision,
  title={Vision-rwkv: Efficient and scalable visual perception with rwkv-like architectures},
  author={Duan, Yuchen and Wang, Weiyun and Chen, Zhe and Zhu, Xizhou and Lu, Lewei and Lu, Tong and Qiao, Yu and Li, Hongsheng and Dai, Jifeng and Wang, Wenhai},
  journal={arXiv preprint arXiv:2403.02308},
  year={2024}
}

@article{liu2024swin,
  title={Swin-UMamba†: Adapting Mamba-based vision foundation models for medical image segmentation},
  author={Liu, Jiarun and Yang, Hao and Zhou, Hong-Yu and Yu, Lequan and Liang, Yong and Yu, Yizhou and Zhang, Shaoting and Zheng, Hairong and Wang, Shanshan},
  journal={IEEE Transactions on Medical Imaging},
  year={2024},
  publisher={IEEE}
}

@inproceedings{codella2018skin,
  title={Skin lesion analysis toward melanoma detection: A challenge at the 2017 international symposium on biomedical imaging (isbi), hosted by the international skin imaging collaboration (isic)},
  author={Codella, Noel CF and Gutman, David and Celebi, M Emre and Helba, Brian and Marchetti, Michael A and Dusza, Stephen W and Kalloo, Aadi and Liopyris, Konstantinos and Mishra, Nabin and Kittler, Harald and others},
  booktitle={2018 IEEE 15th international symposium on biomedical imaging (ISBI 2018)},
  pages={168--172},
  year={2018},
  organization={IEEE}
}

@article{al2020dataset,
  title={Dataset of breast ultrasound images},
  author={Al-Dhabyani, Walid and Gomaa, Mohammed and Khaled, Hussien and Fahmy, Aly},
  journal={Data in brief},
  volume={28},
  pages={104863},
  year={2020},
  publisher={Elsevier}
}

@article{sirinukunwattana2017gland,
  title={Gland segmentation in colon histology images: The glas challenge contest},
  author={Sirinukunwattana, Korsuk and Pluim, Josien PW and Chen, Hao and Qi, Xiaojuan and Heng, Pheng-Ann and Guo, Yun Bo and Wang, Li Yang and Matuszewski, Bogdan J and Bruni, Elia and Sanchez, Urko and others},
  journal={Medical image analysis},
  volume={35},
  pages={489--502},
  year={2017},
  publisher={Elsevier}
}

@inproceedings{jha2019kvasir,
  title={Kvasir-seg: A segmented polyp dataset},
  author={Jha, Debesh and Smedsrud, Pia H and Riegler, Michael A and Halvorsen, P{\aa}l and De Lange, Thomas and Johansen, Dag and Johansen, H{\aa}vard D},
  booktitle={International conference on multimedia modeling},
  pages={451--462},
  year={2019},
  organization={Springer}
}

@inproceedings{deng2009imagenet,
  title={Imagenet: A large-scale hierarchical image database},
  author={Deng, Jia and Dong, Wei and Socher, Richard and Li, Li-Jia and Li, Kai and Fei-Fei, Li},
  booktitle={2009 IEEE conference on computer vision and pattern recognition},
  pages={248--255},
  year={2009},
  organization={Ieee}
}

@inproceedings{milletari2016v,
  title={V-net: Fully convolutional neural networks for volumetric medical image segmentation},
  author={Milletari, Fausto and Navab, Nassir and Ahmadi, Seyed-Ahmad},
  booktitle={2016 fourth international conference on 3D vision (3DV)},
  pages={565--571},
  year={2016},
  organization={Ieee}
}

@article{
zhou2019unet++,
  title={Unet++: Redesigning skip connections to exploit multiscale features in image segmentation},
  author={Zhou, Zongwei and Siddiquee, Md Mahfuzur Rahman and Tajbakhsh, Nima and Liang, Jianming},
  journal={IEEE transactions on medical imaging},
  volume={39},
  number={6},
  pages={1856--1867},
  year={2019},
  publisher={IEEE}
}

@article{oktay2018attention,
  title={Attention u-net: Learning where to look for the pancreas},
  author={Oktay, Ozan and Schlemper, Jo and Folgoc, Loic Le and Lee, Matthew and Heinrich, Mattias and Misawa, Kazunari and Mori, Kensaku and McDonagh, Steven and Hammerla, Nils Y and Kainz, Bernhard and others},
  journal={arXiv preprint arXiv:1804.03999},
  year={2018}
}

@article{chen2021transunet,
  title={Transunet: Transformers make strong encoders for medical image segmentation},
  author={Chen, Jieneng and Lu, Yongyi and Yu, Qihang and Luo, Xiangde and Adeli, Ehsan and Wang, Yan and Lu, Le and Yuille, Alan L and Zhou, Yuyin},
  journal={arXiv preprint arXiv:2102.04306},
  year={2021}
}

@inproceedings{wang2022uctransnet,
  title={Uctransnet: rethinking the skip connections in u-net from a channel-wise perspective with transformer},
  author={Wang, Haonan and Cao, Peng and Wang, Jiaqi and Zaiane, Osmar R},
  booktitle={Proceedings of the AAAI conference on artificial intelligence},
  volume={36},
  number={3},
  pages={2441--2449},
  year={2022}
}

@inproceedings{cao2022swin,
  title={Swin-unet: Unet-like pure transformer for medical image segmentation},
  author={Cao, Hu and Wang, Yueyue and Chen, Joy and Jiang, Dongsheng and Zhang, Xiaopeng and Tian, Qi and Wang, Manning},
  booktitle={European conference on computer vision},
  pages={205--218},
  year={2022},
  organization={Springer}
}

@inproceedings{rahman2024emcad,
  title={Emcad: Efficient multi-scale convolutional attention decoding for medical image segmentation},
  author={Rahman, Md Mostafijur and Munir, Mustafa and Marculescu, Radu},
  booktitle={Proceedings of the IEEE/CVF Conference on Computer Vision and Pattern Recognition},
  pages={11769--11779},
  year={2024}
}

@inproceedings{ibtehaz2023acc,
  title={Acc-unet: A completely convolutional unet model for the 2020s},
  author={Ibtehaz, Nabil and Kihara, Daisuke},
  booktitle={International conference on medical image computing and computer-assisted intervention},
  pages={692--702},
  year={2023},
  organization={Springer}
}

@inproceedings{xu2023mgfuseseg,
  title={Mgfuseseg: Attention-guided multi-granularity fusion for medical image segmentation},
  author={Xu, Guoping and Leng, Xuesong and Li, Chang and He, Xingwei and Wu, Xinglong},
  booktitle={2023 IEEE International Conference on Bioinformatics and Biomedicine (BIBM)},
  pages={3587--3594},
  year={2023},
  organization={IEEE}
}

@inproceedings{valanarasu2022unext,
  title={Unext: Mlp-based rapid medical image segmentation network},
  author={Valanarasu, Jeya Maria Jose and Patel, Vishal M},
  booktitle={International conference on medical image computing and computer-assisted intervention},
  pages={23--33},
  year={2022},
  organization={Springer}
}

@article{ruan2024vm,
  title={Vm-unet: Vision mamba unet for medical image segmentation},
  author={Ruan, Jiacheng and Li, Jincheng and Xiang, Suncheng},
  journal={arXiv preprint arXiv:2402.02491},
  year={2024}
}

@article{huang2021missformer,
  title={Missformer: An effective medical image segmentation transformer},
  author={Huang, Xiaohong and Deng, Zhifang and Li, Dandan and Yuan, Xueguang},
  journal={arXiv preprint arXiv:2109.07162},
  year={2021}
}

@article{wu2403h,
  title={H-vmunet: High-order vision mamba unet for medical image segmentation. arXiv 2024},
  author={Wu, R and Liu, Y and Liang, P and Chang, Q},
  journal={arXiv preprint arXiv:2403.13642}
}

@inproceedings{he2024frcnet,
  title={FRCNet: Frequency and region consistency for semi-supervised medical image segmentation},
  author={He, Along and Li, Tao and Wu, Yanlin and Zou, Ke and Fu, Huazhu},
  booktitle={International Conference on Medical Image Computing and Computer-Assisted Intervention},
  pages={305--315},
  year={2024},
  organization={Springer}
}

@article{vaswani2017attention,
  title={Attention is all you need},
  author={Vaswani, Ashish and Shazeer, Noam and Parmar, Niki and Uszkoreit, Jakob and Jones, Llion and Gomez, Aidan N and Kaiser, {\L}ukasz and Polosukhin, Illia},
  journal={Advances in neural information processing systems},
  volume={30},
  year={2017}
}

@article{dosovitskiy2020image,
  title={An image is worth 16x16 words: Transformers for image recognition at scale},
  author={Dosovitskiy, Alexey and Beyer, Lucas and Kolesnikov, Alexander and Weissenborn, Dirk and Zhai, Xiaohua and Unterthiner, Thomas and Dehghani, Mostafa and Minderer, Matthias and Heigold, Georg and Gelly, Sylvain and others},
  journal={arXiv preprint arXiv:2010.11929},
  year={2020}
}

@article{gu2023mamba,
  title={Mamba: Linear-time sequence modeling with selective state spaces},
  author={Gu, Albert and Dao, Tri},
  journal={arXiv preprint arXiv:2312.00752},
  year={2023}
}

@article{liu2024vmamba,
  title={Vmamba: Visual state space model},
  author={Liu, Yue and Tian, Yunjie and Zhao, Yuzhong and Yu, Hongtian and Xie, Lingxi and Wang, Yaowei and Ye, Qixiang and Jiao, Jianbin and Liu, Yunfan},
  journal={Advances in neural information processing systems},
  volume={37},
  pages={103031--103063},
  year={2024}
}

@inproceedings{gong2025nnmamba,
  title={nnmamba: 3d biomedical image segmentation, classification and landmark detection with state space model},
  author={Gong, Haifan and Kang, Luoyao and Wang, Yitao and Wang, Yihan and Wan, Xiang and Wu, Xusheng and Li, Haofeng},
  booktitle={2025 IEEE 22nd International Symposium on Biomedical Imaging (ISBI)},
  pages={1--5},
  year={2025},
  organization={IEEE}
}

@article{he2022progressive,
  title={Progressive multiscale consistent network for multiclass fundus lesion segmentation},
  author={He, Along and Wang, Kai and Li, Tao and Bo, Wang and Kang, Hong and Fu, Huazhu},
  journal={IEEE transactions on medical imaging},
  volume={41},
  number={11},
  pages={3146--3157},
  year={2022},
  publisher={IEEE}
}

@article{yuan2024mamba,
  title={Mamba or rwkv: Exploring high-quality and high-efficiency segment anything model},
  author={Yuan, Haobo and Li, Xiangtai and Qi, Lu and Zhang, Tao and Yang, Ming-Hsuan and Yan, Shuicheng and Loy, Chen Change},
  journal={arXiv preprint arXiv:2406.19369},
  year={2024}
}

@article{zhai2021attention,
  title={An attention free transformer},
  author={Zhai, Shuangfei and Talbott, Walter and Srivastava, Nitish and Huang, Chen and Goh, Hanlin and Zhang, Ruixiang and Susskind, Josh},
  journal={arXiv preprint arXiv:2105.14103},
  year={2021}
}

@inproceedings{zhou2024spatial,
  title={Spatial-frequency dual domain attention network for medical image segmentation},
  author={Zhou, Zhenhuan and He, Along and Wu, Yanlin and Yao, Rui and Xie, Xueshuo and Li, Tao},
  booktitle={2024 IEEE International Conference on Bioinformatics and Biomedicine (BIBM)},
  pages={4076--4081},
  year={2024},
  organization={IEEE}
}

@article{peng2023rwkv,
  title={Rwkv: Reinventing rnns for the transformer era},
  author={Peng, Bo and Alcaide, Eric and Anthony, Quentin and Albalak, Alon and Arcadinho, Samuel and Biderman, Stella and Cao, Huanqi and Cheng, Xin and Chung, Michael and Grella, Matteo and others},
  journal={arXiv preprint arXiv:2305.13048},
  year={2023}
}

@ARTICLE{11039697,
  author={Xu, Guoxia and Liu, Yang and Deng, Lizhen and Wang, Xiaokang and Zhu, Hu},
  journal={IEEE Transactions on Aerospace and Electronic Systems}, 
  title={SMNet: A Semantic Guided Mamba Network for Remote Sensing Change Detection}, 
  year={2025},
  volume={},
  number={},
  pages={1-12},
  doi={10.1109/TAES.2025.3580691}}

@inproceedings{xu2025urwkv,
  title={URWKV: Unified RWKV Model with Multi-state Perspective for Low-light Image Restoration},
  author={Xu, Rui and Niu, Yuzhen and Li, Yuezhou and Xu, Huangbiao and Liu, Wenxi and Chen, Yuzhong},
  booktitle={Proceedings of the Computer Vision and Pattern Recognition Conference},
  pages={21267--21276},
  year={2025}
}

@article{zhou2025personalizable,
  title={Personalizable Long-Context Symbolic Music Infilling with MIDI-RWKV},
  author={Zhou-Zheng, Christian and Pasquier, Philippe},
  journal={arXiv preprint arXiv:2506.13001},
  year={2025}
}

@article{zhang2025out,
  title={Out-of-Distribution Semantic Occupancy Prediction},
  author={Zhang, Yuheng and Duan, Mengfei and Peng, Kunyu and Wang, Yuhang and Liu, Ruiping and Teng, Fei and Luo, Kai and Li, Zhiyong and Yang, Kailun},
  journal={arXiv preprint arXiv:2506.21185},
  year={2025}
}

@inproceedings{cciccek20163d,
  title={3D U-Net: learning dense volumetric segmentation from sparse annotation},
  author={{\c{C}}i{\c{c}}ek, {\"O}zg{\"u}n and Abdulkadir, Ahmed and Lienkamp, Soeren S and Brox, Thomas and Ronneberger, Olaf},
  booktitle={International conference on medical image computing and computer-assisted intervention},
  pages={424--432},
  year={2016},
  organization={Springer}
}

@article{chen2022aau,
  title={AAU-net: an adaptive attention U-net for breast lesions segmentation in ultrasound images},
  author={Chen, Gongping and Li, Lei and Dai, Yu and Zhang, Jianxun and Yap, Moi Hoon},
  journal={IEEE Transactions on Medical Imaging},
  volume={42},
  number={5},
  pages={1289--1300},
  year={2022},
  publisher={IEEE}
}

@article{li2023can,
  title={CAN: Context-assisted full Attention Network for brain tissue segmentation},
  author={Li, Zhan and Zhang, Chunxia and Zhang, Yongqin and Wang, Xiaofeng and Ma, Xiaolong and Zhang, Hai and Wu, Songdi},
  journal={Medical Image Analysis},
  volume={85},
  pages={102710},
  year={2023},
  publisher={Elsevier}
}

@article{li2018h,
  title={H-DenseUNet: hybrid densely connected UNet for liver and tumor segmentation from CT volumes},
  author={Li, Xiaomeng and Chen, Hao and Qi, Xiaojuan and Dou, Qi and Fu, Chi-Wing and Heng, Pheng-Ann},
  journal={IEEE transactions on medical imaging},
  volume={37},
  number={12},
  pages={2663--2674},
  year={2018},
  publisher={IEEE}
}

@article{bui20173d,
  title={3D densely convolutional networks for volumetric segmentation},
  author={Bui, Toan Duc and Shin, Jitae and Moon, Taesup},
  journal={arXiv preprint arXiv:1709.03199},
  year={2017}
}

@inproceedings{hatamizadeh2022unetr,
  title={Unetr: Transformers for 3d medical image segmentation},
  author={Hatamizadeh, Ali and Tang, Yucheng and Nath, Vishwesh and Yang, Dong and Myronenko, Andriy and Landman, Bennett and Roth, Holger R and Xu, Daguang},
  booktitle={Proceedings of the IEEE/CVF winter conference on applications of computer vision},
  pages={574--584},
  year={2022}
}

@article{he2023h2former,
  title={H2former: An efficient hierarchical hybrid transformer for medical image segmentation},
  author={He, Along and Wang, Kai and Li, Tao and Du, Chengkun and Xia, Shuang and Fu, Huazhu},
  journal={IEEE Transactions on Medical Imaging},
  volume={42},
  number={9},
  pages={2763--2775},
  year={2023},
  publisher={IEEE}
}

@article{liao2024lightm,
  title={Lightm-unet: Mamba assists in lightweight unet for medical image segmentation},
  author={Liao, Weibin and Zhu, Yinghao and Wang, Xinyuan and Pan, Chengwei and Wang, Yasha and Ma, Liantao},
  journal={arXiv preprint arXiv:2403.05246},
  year={2024}
}

@article{wang2024large,
  title={Large window-based mamba unet for medical image segmentation: Beyond convolution and self-attention},
  author={Wang, Jinhong and Chen, Jintai and Chen, Danny and Wu, Jian},
  journal={CoRR},
  year={2024}
}

@article{wang2025feat,
  title={FEAT: Full-Dimensional Efficient Attention Transformer for Medical Video Generation},
  author={Wang, Huihan and Yang, Zhiwen and Zhang, Hui and Zhao, Dan and Wei, Bingzheng and Xu, Yan},
  journal={arXiv preprint arXiv:2506.04956},
  year={2025}
}

@inproceedings{he2025rwkvmatch,
  title={RWKVMatch: Vision RWKV-based Multi-scale Feature Matching Network for Unsupervised Deformable Medical Image Registration},
  author={He, Zixuan and Tang, Jing and Zhao, Zitong and Gong, Zeyu},
  booktitle={ICASSP 2025-2025 IEEE International Conference on Acoustics, Speech and Signal Processing (ICASSP)},
  pages={1--5},
  year={2025},
  organization={IEEE}
}

@inproceedings{farshad2022net,
  title={Y-net: A spatiospectral dual-encoder network for medical image segmentation},
  author={Farshad, Azade and Yeganeh, Yousef and Gehlbach, Peter and Navab, Nassir},
  booktitle={International conference on medical image computing and computer-assisted intervention},
  pages={582--592},
  year={2022},
  organization={Springer}
}

@inproceedings{li2023mixunet,
  title={Mixunet: Mix the 2d and 3d models for robust medical image segmentation},
  author={Li, Jiawei and Chen, Shizhan and Ma, Shiqiang and Guo, Fei and Tang, Jijun},
  booktitle={2023 IEEE International Conference on Bioinformatics and Biomedicine (BIBM)},
  pages={1242--1247},
  year={2023},
  organization={IEEE}
}

@inproceedings{huang2023polyp2former,
  title={Polyp2Former: Boundary Guided Network Based on Transformer for Polyp Segmentation},
  author={Huang, Xiaoshuang and Huang, Jinze and Wang, Shuo and Wei, Yaoguang and An, Dong and Liu, Jincun},
  booktitle={2023 IEEE International Conference on Bioinformatics and Biomedicine (BIBM)},
  pages={1971--1976},
  year={2023},
  organization={IEEE}
}

@article{wang2025set,
  title={SET: Superpixel Embedded Transformer for skin lesion segmentation},
  author={Wang, Zhonghua and Lyu, Junyan and Tang, Xiaoying},
  journal={Medical Image Analysis},
  pages={103738},
  year={2025},
  publisher={Elsevier}
}

@INPROCEEDINGS{10822736,
  author={Zheng, Fuchen and Chen, Xuhang and Liu, Weihuang and Li, Haolun and Lei, Yingtie and He, Jiahui and Pun, Chi-Man and Zhou, Shoujun},
  booktitle={2024 IEEE International Conference on Bioinformatics and Biomedicine (BIBM)}, 
  title={SMAFormer: Synergistic Multi-Attention Transformer for Medical Image Segmentation}, 
  year={2024},
  volume={},
  number={},
  pages={4048-4053},
  doi={10.1109/BIBM62325.2024.10822736}}

@INPROCEEDINGS{10821714,
  author={Qiao, Sibo and Zhao, Zhiyuan and Xie, Pengfei and Yin, Wenjing and Wang, Min and Zhang, Yuanyuan and Pang, Shanchen},
  booktitle={2024 IEEE International Conference on Bioinformatics and Biomedicine (BIBM)}, 
  title={A Novel Conv-Mamba-Hybrid Network for Medical Image Segmentation}, 
  year={2024},
  volume={},
  number={},
  pages={4261-4268},
  doi={10.1109/BIBM62325.2024.10821714}}

@INPROCEEDINGS{10821761,
  author={Chen, Chaowei and Yu, Li and Min, Shiquan and Wang, Shunfang},
  booktitle={2024 IEEE International Conference on Bioinformatics and Biomedicine (BIBM)}, 
  title={MSVM-UNet: Multi-Scale Vision Mamba UNet for Medical Image Segmentation}, 
  year={2024},
  volume={},
  number={},
  pages={3111-3114},
  doi={10.1109/BIBM62325.2024.10821761}}

@inproceedings{ghazali2007feature,
  title={Feature extraction technique using discrete wavelet transform for image classification},
  author={Ghazali, Kamarul Hawari and Mansor, Mohd Fais and Mustafa, Mohd Marzuki and Hussain, Aini},
  booktitle={2007 5th Student Conference on Research and Development},
  pages={1--4},
  year={2007},
  organization={IEEE}
}

@inproceedings{zhou2023xnet,
  title={Xnet: Wavelet-based low and high frequency fusion networks for fully-and semi-supervised semantic segmentation of biomedical images},
  author={Zhou, Yanfeng and Huang, Jiaxing and Wang, Chenlong and Song, Le and Yang, Ge},
  booktitle={Proceedings of the IEEE/CVF international conference on computer vision},
  pages={21085--21096},
  year={2023}
}

@inproceedings{huang2025wavelet,
  title={Wavelet-Assisted Multi-Frequency Attention Network for Pansharpening},
  author={Huang, Jie and Huang, Rui and Xu, Jinghao and Peng, Siran and Duan, Yule and Deng, Liang-Jian},
  booktitle={Proceedings of the AAAI Conference on Artificial Intelligence},
  volume={39},
  number={4},
  pages={3662--3670},
  year={2025}
}
\end{document}